\magnification=\magstep2
\baselineskip=14.4pt
\parindent=15pt

\centerline{\bf QBism in the New Scientist}

\vskip 10pt

\centerline {N. David Mermin}

\centerline{Laboratory of Atomic and Solid State Physics}

\centerline{Cornell University, Ithaca, NY 14853}
\vskip 20pt

{\narrower \narrower

I correct a misrepresentation of QBism as antirealist.

}

\vskip 20pt 

On 10 May 2014 the New Scientist published an article that repeated many naive misunderstandings  of QBism.   Since several of these misrepresentations are attributed to me, I submitted the following letter to the editor: 

\vskip 10pt

{\narrower

I am delighted that you take the QBist understanding of science seriously enough to feature it on your cover (State of Mind, 32-35, 10 May 2014).   But the headline on the cover, the title of the article, and several statements within it all overemphasize the subjectivity of the scientist almost as much as conventional physics underemphasizes it by ignoring it entirely.    QBism strives to balance the subjective and the objective.

For example your article attributes to QBism the view that ``Measurements do not cause things to happen in the real world, whatever that is: they cause things to happen in our heads.''  The actual QBist position is that a measurement is any action a particular person (Alice) takes on her external world, and the outcome of the measurement is the experience this world induces back in Alice through its response to her action.  This differs from your formulation in several ways:

Just as important as the action of the scientist on the world is the response of the world to that action.     Alice does not doubt the existence of this world.   What happens only in Alice's head is what quantum theory calls the ``outcome" of the measurement.   Other consequences of Alice's action, though not as immediately accessible to her as its outcome, are part of her external world and potentially accessible to others, through their own actions.    Alice,  like all users of the quantum theory, has her own private subjective experience, but she can attempt to describe this to others through the imperfect medium of language; this helps to account for the common features of the different external worlds that each of us individually infers from our own private experience.

It is, of course, hard to convey all this in three pages, a few headlines, and a very short letter to the editor.   It has, after all, escaped the awareness of almost all physicists for nearly 90 years.   For a more nuanced view of QBism I recommend the paper cited in your article, posted as arXiv:1311.5253.  

} 

\vskip 10pt

I promptly received an email pointing out that letters should not exceed 250 words, and that ``Our deadlines do not leave time to consult you on the editing of letters. We're quite good at saying what you meant, though.''   I
immediately withdrew the above letter and resubmitted an abbreviated text:

\vskip 10pt
{\narrower

I am delighted that you take the QBist understanding of science seriously (State of Mind, 32-35, 10 May 2014).   But  you overemphasize the subjectivity of the scientist as badly as conventional physics ignores it.

You attribute to QBism the view that ``Measurements do not cause things to happen in the real world, whatever that is: they cause things to happen in our heads.'' Actually QBists take a measurement as any action anybody (Alice) takes on the world, and the outcome of the measurement is the experience the world induces back in Alice through its response to her action.

\vfil\eject

Just as important as the action of Alice on the world is the response of the world to her action.     Alice does not doubt the existence of the world.   What happens only in Alice's head is the ``outcome" of the measurement.   Other consequences of Alice's action, though not as  accessible to her as its outcome, belong to the world and are accessible to others, through their own actions.    Alice has her private subjective experience, but can attempt to describe this to others with language; this helps to account for the common features of the different external worlds that each of us  infers from our private experience.

It is hard to convey all this in three pages, a few headlines, and a very short letter to the editor.   It has, after all, escaped the awareness of almost all physicists for nearly 90 years 

}

\vskip 10pt

On 7 June 2014 the New Scientist (p.~30) published a letter drawn from both of my versions.   The letter they published differs from both of mine in  two important ways:

1.   The phrase ``whatever that is" is no longer in my quotation from the article,
though it is there in both  versions of my letter.

2.   The first three sentences of my next to last paragraph,  which are quite similar in both versions, are gone.  (My second version of that paragraph  is at the top of this page.)

The first omission --- of the New Scientist's own words ---  diminishes the degree to which their  article misrepresents QBism as antirealist.      The second omission --- from both my versions ---   eliminates the heart of my explanation of QBist realism.   Their combined effect is to turn my correction of the New Scientist's  gross misrepresentation of realism in QBism  into what sounds like a pedantic quibble. 

\bye